\documentstyle[12pt]{article}
\newcommand{\esp}{\vspace{.1cm}}
\newcommand{\ba}{\begin{array}}
\newcommand{\ea}{\end{array}}
\newcommand{\bc}{\begin{center}}
\newcommand{\ec}{\end{center}}
\newcommand{\be}{\begin{equation}}
\newcommand{\ee}{\end{equation}}
\newcommand{\bea}{\begin{eqnarray}}
\newcommand{\eea}{\end{eqnarray}}
\newcommand{\bdm}{\begin{displaymath}}
\newcommand{\edm}{\end{displaymath}}
\newcommand{\plb}[1]{Phys. Lett. {\bf #1}}

\newcommand{\prb}[1]{Phys. Rev. {\bf #1}}
\newcommand{\apb}[1]{Ann. Phys. (N.Y.) {\bf #1} }

\begin{document}
\begin{center}
\begin{Large}
\begin{bf}
 Potential Harmonic Calculations of \\
 Helium Triplet States \\
\end{bf}
\end{Large}
\vspace{.7cm}
Anthony  D. Klemm \\
School of Computing \& Mathematics, Deakin University, Geelong, Australia \\
\esp
 Michel Fabre de la Ripelle \\
        Institut de Physique Nucl\'eaire, 91406 Orsay Cedex, France \\
{\it and} \\
Sigurd Yves Larsen \\
Department of Physics, Temple University, Philadelphia, PA 19122, USA  \\
Physics Department, University of South Africa 0003, Pretoria 0003, 
South Africa \\
\vspace{1.cm}
\begin{bf}
ABSTRACT \\
\end{bf}
\vspace{.3cm}
\end{center}

The hyperspherical harmonics (HH) provide a complete basis for the
expansion of atomic wave functions, but even for two particles the number of
harmonics for a given order is not trivial and, as the number of electrons
increases, this degeneracy becomes absolutely prohibitive. 
We modify the method by selecting a subset of the basis that, we feel, will
yield the physically most important part of the wave function, and test the
idea on simple systems.

\esp
 In a previous work (with M. Haftel) of the ground and first
excited ${}^1S_0$ states of the helium atom we found that the error in the
binding energy of the ${}^1S_0$ ground state was of the order of $1$ part
in $10^{4}$ and 
that it decreased for the first excited state.

\esp
We now have applied our method to the equivalent triplet ${}^3S_1$ states.
We report on this work, and our results, and hope to draw attention to the
interesting accuracy that we obtain with the relatively simple wave functions
of our formulation. 

\newpage
\topmargin=0in
\pagestyle{plain}
\pagenumbering{arabic}
\setcounter{page}{2}
\vspace{0.6cm}
\begin{center}
\section*{Introduction}
\end{center}

 The bound states of the helium atom have been studied by using several
methods.

The most accurate results have been achieved by variational calculations
where the states are described by wave functions which include more than
$200$ parameters \cite{Frank}. Other, more recent, calculations have been
performed using expansions in hyperspherical harmonics (HH), where
the convergence was speeded up by the introduction of a correlation
factor describing the average behaviour of the wave function at short
and long ranges\cite{Man}. The residual part of the wave function was
expanded in HH. This leads to an accuracy of about nine figures in the
binding energy, but with a nonuniform convergence as the number of HH
is increased.

 In our approach to the many body problem, we select the HH occurring in
the wave function by following a scheme in which the basis elements are
introduced according to the degree of complexity of the correlations
that they describe.

In the first approximation, a HH of minimal degree (because of the Pauli
principle) defines the state. In the next approximation, terms will appear
in our equations consisting of the product of the HH of minimal degree and
the potential. The basis elements that can correctly
reproduce these terms form our Potential Harmonic basis\cite{fab}, which we
use in this work.
A further approximation would be obtained by applying the potential
to the wave function, including the correlations already obtained from
the previous order and by taking into account the new basis elements
generated by this product.

 In the analysis of the ground and first excited ${}^1S_0$ state of the
helium atom \cite{haf}, what was studied was the accuracy obtained
by using the PH basis.
The resulting expansion of the wave function provides a simple
structure where one and two body contributions are exactly separated.
It leads to a simple description of the wave function.

We found that the error in our calculation of 
the binding energy of the ${}^1S_0$ ground state was of the order
of $10^{-4}$ and decreased for the first excited state.

Our purpose, then, in this paper was to study the worth of the method for
triplet ${}^3S_1$ states. Since the potential basis is
complete in so far as describing the independent motion of the
electrons (i.e. when the interaction between the electrons is neglected),
the missing energy must come from the Coulomb interaction between the
two electrons.
In the ${}^1S_0$ state this interaction is maximum at short distances
and is expected to be larger in the space symmetric ${}^1S_0$ states
than in the space anti-symmetric ${}^3S_1$ states, where the effect of
this interaction damps out at small distances where the wave function
goes to zero.
We were therefore very hopeful that our method would be even more satisfactory
for the triplet than for the singlet.
\newpage
\begin{center}
\section*{Formalism}
\end{center}

 In the HH expansion method the coordinates are expressed in polar
form in a $D$-dimensional space where $D = 3N$ is the number of
independent space degrees of freedom. $N=2$ for two electrons with position
vectors $\vec{x}_1$ and $\vec{x}_2$, relative to the nucleus with
nuclear charge $Z$ and polar coordinates $(r_i, \vec{\omega}_i),\;i=1,2$, 
where the $\vec{\omega}_i$ denote the two standard angle measures for the
$i$th electron,
the hyperspherical coordinates $(r, \Omega)$ are chosen to be
the hyperradius $r = \sqrt{r_1^2 + r_2^2}$ and the set $\Omega$ of five
angular coordinates, $\vec{\omega}_1$, $\vec{\omega}_2$ and $z$, where 
 $r_1 = r\cos \phi$, $r_2 = r\sin \phi$ and we replace $\phi$ by
\be z = \cos 2\phi = \frac{r_1^2 - r_2^2}{r^2} \label{0} \ee

The Schr\"odinger equation for the helium atom, in atomic units
$(\hbar = m = e = 1)$,
\be
\left\{ -\frac{1}{2}(\nabla^2_1 + \nabla^2_2) - Z(\frac{1}{r_1}
 + \frac{1}{r_2}) + \frac{1}{r_{12}}
  - E\right\}\,\Psi = 0\,, \label{1}
\ee
is then transformed, in hyperspherical coordinates, into
\bea
\left\{ - \frac{1}{2}\left[\frac{\partial^2}{\partial r^2} +
 \frac{L^2(\Omega)}{r^2}\right] \right.
& - & \frac{\sqrt{2}Z}{r}\left(\frac{1}{\sqrt{1+z}} \right.  +  \nonumber \\
 \left. \frac{1}{\sqrt{1-z}}\right)
& + & \left.  \frac{1}{r}\frac{1}{\sqrt{1 - x\sqrt{1 - z^2}}} -
E\right\}\,\Phi(r,\Omega) =  0 \label{2}
\eea
 where $x$ is the cosine of the angle between the two electrons,
($x = \vec{x}_1.\vec{x}_2/(r_1 r_2) = \cos \theta$),
 $L^2(\Omega)$ is the grand orbital operator,  and
 $\Psi$ and $\Phi$ are related by
\be \Psi = r^{-5/2}\Phi(r,\Omega)\,. \label{3} \ee
We note that if $Y_{[L]}(\Omega)$ is a HH, with  $r^LY_{[L]}(\Omega)$ the
associated harmonic polynomial of degree $L$, then $L^{2}(\Omega)$ in
equation (\ref{2}) satisfies 
\be [L^2(\Omega) + {\cal L}({\cal L} + 1)] Y_{[L]}(\Omega) = 0 \label{4}
\ee
with ${\cal L} = L + 3/2$ in the six
 dimensional space spanned by the
coordinates $\vec{x}_1$ and $\vec{x}_2$. Also, the Hamiltonian in
(\ref{2}) is an even function of $z$  and therefore the wave function can
have a definite $z$-parity. The $z$-parity is associated, according to
(\ref{0}), with the parity in the exchange of the two electrons, i.e. it
is even for the singlet ${}^1S_0$ states and odd for the triplet ${}^3S_1$
states.

When the interaction between the electrons is set to zero, equation (\ref{2})
is separable and the wave function for $S$-states can be written either as
a product of individual wave functions
\be
\Psi = \phi_{n_1l_1}(r_1)\;\phi_{n_2l_2}(r_2) \pm
     \phi_{n_2l_2}(r_1)\;\phi_{n_1l_1}(r_2) \label{5} \ee
 where $+(-)$ is for space (anti)symmetric wave functions, or as a function
\be \Psi = \frac{\phi(z,r)}{r^{5/2}} \label{6} \ee
with a $z$-parity even(odd) for space (anti)symmetric wave functions.
Both descriptions are equivalent.
\newpage

Let the HH expansion of the wave function in (\ref{2}) be
\be \Phi(r,\Omega) = \sum_{[L]} Y_{[L]}(\Omega)\,u_{[L]}(r) \label{7}
\ee
where $[L]$ is the set of the five quantum numbers $[L,l_1,l_2,m_1,m_2]$
associated with the five degrees of freedom $(\vec{\omega}_1, \vec{\omega}_2,
z)$. The $(l_i,m_i)$ are the orbital quantum numbers and
$L = 2k + l_1 + l_2$  where $k$ is the quantum number
associated with $z$.

The coupled equations for the partial waves $u_{[L]}(r)$ are
\be \left\{ \frac{1}{2}\left[ -\frac{d^2}{dr^2} + \frac{{\cal L}(
 {\cal L} + 1)}{r^2}\right] - E \right\}\,u_{[L]}(r) \\
 + \sum_{[L']} <Y_{[L]} | V(r, \Omega) | Y_{[L']}>\,u_{[L']}(r) = 0
 \label{8} \ee
where in our problem $V(r,\Omega)$ is the potential occurring in equation
 (\ref{2}).

In equation (\ref{7}) the sum starts from a minimal degree $L_m$. In the
hypercentral approximation (HCA) only the first term in equation (\ref{7})
is taken into account and equation (\ref{8}) is reduced to a
 hyperradial equation
\be \left\{ \frac{1}{2} \left[ - \frac{d^2}{dr^2} + 
 \frac{{\cal L}_m({\cal L}_m + 1)}{r^2}\right] + U_{[L_m]}(r)
 - E \right\}\, u_{[L_m]}(r) = 0\,, \label{9} \ee
where
\be U_{[L_m]}(r) = <Y_{[L_m]} | V(r,\Omega) | Y_{[L_m]}> \,. \label{10} \ee
For a normalized HH
\[ \int \left| Y_{[L_m]}(\Omega)\right|^2\; d\Omega = 1\,. \]
Here, the surface element of the unit hypersphere is
\begin{eqnarray*}
 d\Omega & = &
d\vec{\omega}_1\,d\vec{\omega}_2\;(\sin \phi\,\cos\phi)^2\;d\phi \nonumber \\
 & = & d\vec{\omega}_1\,d\vec{\omega}_2\,\frac{1}{8}\sqrt{1-z^2}\,dz\,.
\end{eqnarray*}
For $S$-states
\be Y_{[2k,0,0,0,0]}(\Omega) = \pi^{-3/2}C_k^1(z) \label{11} \ee
and
\[ d\Omega  =  \pi^2\,\sin \theta\,d\theta\,\sqrt{1-z^2}\,dz\,. \]
With the Gegenbauer polynomials $C_0^1 = 1$ and $C_1^1(z) = 2z$, the
hypercentral potential
for the ${}^1S_0$ and ${}^3S_1$ ground states are, for $k=0,\;1$ respectively,
\be U_{[0]}(r) = \frac{1}{\pi}\int V(r,\Omega)\sin\theta\,d\theta
 \sqrt{1-z^2}\,dz \label{12} \ee
and
\be U_{[2]}(r) = \frac{4}{\pi} \int V(r,\Omega)\sin\theta\,d\theta
 \,z^2\,\sqrt{1-z^2}\,dz \label{13} \ee
where the integrals are taken in the domain $0 \le \theta \le \pi$ and
 $-1 \le z \le 1$.
\newpage

For a Coulomb potential, the integrals (\ref{12}) and (\ref{13}) are
simply proportional to $1/r$, and therefore the HCA equation (\ref{9})
can be solved analytically \cite{fab}.
In order
to go further, another variable must be introduced. An expansion 
of $r_{12}^{-1}$ in terms of
Legendre polynomials is obtained from the generating function 
\be \sum_{\ell=0}^\infty P_\ell(u)\,z^\ell = (1 - 2zu + z^2)^{-1/2} \, ,
\label{14} \ee
leading to
\begin{eqnarray}
\frac{1}{r_{12}} & = & \frac{1}{r} \frac{1}{\sqrt{1-\sin2\phi\,\cos\theta}}
\nonumber \\
 & = & \frac{1}{r}\sqrt{\frac{2}{1+|z|}}\sum_{\ell=0}^\infty
\left(\frac{1-|z|}{1+|z|}\right)^{\ell/2}P_\ell(x) \, .
\label{15} \end{eqnarray}
We expand the wave function for electrons in $S$-states in terms
of normalized Legendre polynomials:
\be \Phi(r,z,x) = \sum_{\ell=0}^\infty \overline{P}_\ell(x)
  \;u_\ell(z,r)\,, \label{16} \ee
where $\overline{P}_(x) = \sqrt{\ell+1/2}\,P_\ell(x)$.
Introducing this form for the wave function into Eq. (\ref{2}),
we obtain an infinite system of coupled second order partial
differential equations for the partial waves $u_\ell(z,r)$:
\bea
\left\{ \frac{\partial^2}{\partial r^2} \right.& + & \left.
 \frac{{L}^2(l,z)}{r^2}
 + 2\sqrt{2}\,\frac{Z}{r}\left[ \frac{1}{\sqrt{1+z}} + \frac{1}{\sqrt{1-z}}
 \right] + 2E \right\}\;u_\ell(z,r) \nonumber \\
 & = & 2 \sum_{\ell'=0}^\infty <\overline{P}_\ell | \frac{1}{r_{12}} |
 \overline{P}_{\ell'} > u_{\ell'}(z,r)   \label{17} 
\eea
where
\be L^{2}(l,z) = -\frac{15}{4}\frac{1}{r^2}
      + \frac{4}{\sqrt{1 - z^{2}}}\frac{\partial}{\partial z}
  (1-z^{2})^{3/2} \frac{\partial}{\partial z} - \frac{4}{1-z^{2}}\, 
   \ell(\ell +1)
 \ee
and the matrix element
\bdm 
\left<\overline{P}_\ell \mid \frac{1}{r_{12}} \mid \overline{P}_{\ell'}
 \right> =
\sqrt{\left(\ell+\frac{1}{2}\right)\left(\ell'+\frac{1}{2}\right)} \int_{-1}^1
 P_\ell(x)\frac{1}{r_{12}} P_{\ell'}(x)\,dx  
\edm
where $r_{12}^{-1}$ is given by (\ref{15}) in terms of $r$, $z$ and
$x = \cos\theta$. 
Rather than try to solve this system of coupled two-variable
second order partial differential equations to atomic accuracy, we
expand $\Phi(r,z,x)$ in hyperspherical harmonics. To then solve exactly
equations (\ref{17}) requires the complete Zernike-Brinkman basis \cite{zern}
HE for two particles in $S$-states, i.e. a basis complete in the 
variable $z$ for each value of $\ell$ in the expansion (\ref{16}). This
would lead us to a 
large number of coupled equations (\ref{8}) to be solved for a given
accuracy and we try to reduce the size of the basis and choose the potential
basis.

This
basis consists of: (i) the basis for $\ell=0$
which is complete for an expansion of any function of $z$ only (in
particular of the potential $r_1^{-1} + r_2^{-1}$), and (ii) the
basis needed for a complete expansion of $r_{12}^{-1}$.
This implies that equation (\ref{2}) can then be solved with a harmonic 
expansion which includes
only two orthogonal elements for each grand orbital $L=2k$, except 
when $L = 0$.

The PH approximation selects the partial waves in expansion (\ref{7})
which are coupled to the largest components, $L=0$ ($L=2$), of the
${}^1S_0$ (${}^3S_1$) ground states of the wave function respectively.
We call this the optimal subset.

In order to extract this subset, we introduce the
kinematic rotation vector
 $\vec{R}(\varphi) = \vec{r}_1\cos \varphi + \vec{r}_2 \sin \varphi$,
with parameter $\varphi$,
where $\vec{R}(0) = \vec{r}_1$, $\vec{R}(\pi/2)=\vec{r}_2$,
$\vec{R}(3\pi/4)=(\vec{r}_2 - \vec{r}_1)/\sqrt{2}$, and we define
$z(\varphi) = 2 R^2(\varphi)/r^2 - 1$, with $z(0) = \cos 2\phi=z$,
leading to $z(\varphi) = z \cos 2\varphi + x\sqrt{1-z^2}\sin 2\varphi$
and
\be
\frac{1}{r_1} = \frac{\sqrt{2}}{\sqrt{1+z}}\,,\ \ \frac{1}{r_2} =
\frac{\sqrt{2}}{\sqrt{1-z}}\,,\ \ \frac{1}{r_{12}}  = \frac{1}
{r\sqrt{1+ z(3\pi/4)}} \label{20}
\ee
 for $\varphi=0$, $\pi/2$, and $3\pi/4$, respectively.

The Potential Harmonics for the $S$-state are given in terms of the
kinematic angle $\varphi$ by
\be {\cal P}_{2k}^0(\Omega,\varphi) = \pi^{-3/2} C_k^1
(z(\varphi)) \label{21} \ee
where $C_k^1$ is a Gegenbauer polynomial of even grand orbital
$L = 2k$. We use the notation ${\cal P}_{2k}^0(\Omega_i)$,
$i = 1,\ 2,\ 12$, for ${\cal P}_{2k}^0(\Omega,\varphi_i)$,
where $\varphi=0,\ \pi/2,\ 3\pi/4$ refer to the 
respective choices of the kinetic angle parameter for each coordinate.
The PH
expansion of $r_i^{-1}$, for $i=1,\ 2,\ 12$, is then given in \cite{fab}
for $n=-1$ by
\[ \frac{1}{r_i} = \frac{16\sqrt{\pi}}{r} \sum_k (-1)^k \frac{k+1}
{(2k+1)(2k+3)} {\cal P}_{2k}^0(\Omega_i)\,,\ i=1,2 \]
where ${\cal P}_{2k}^0(\Omega_2) = (-1)^k {\cal P}_{2k}^0(\Omega_1)$,
according to the parity of the Gegenbauer polynomial, and
\[ \frac{1}{r_{12}} = \frac{8\sqrt{2\pi}}{r} \sum_k (-1)^k
\frac{k+1}{(2k+1)(2k+3)}{\cal P}_{2k}^0(\Omega_{12})\,. \]

In a PH expansion of the wave function for the singlet $S$-state, the 
optimal subset consists of: (i) ${\cal P}_{2k}^0(\Omega_1)$ for
$k$ even and (ii) that part of ${\cal P}_{2k}^0(\Omega_{12})$ orthogonal
to ${\cal P}_{2k}^0(\Omega_1)$. One uses the orthonormalized base
$B_k(\Omega) = {\cal P}_{2k}^0(\Omega_1)$ for $k$ even,
$B_k^\bot(\Omega) = {\cal P}_{2k}^0(\Omega_{12})$ for $k$ odd, and
$B_k^\bot(\Omega) = \frac{(k+1)}{\sqrt{k(k+2)}}\left[
{\cal P}_{2k}^0(\Omega_{12}) - \frac{1}{(k+1)}{\cal P}_{2k}^0
 (\Omega_1)\right]$ for $k$ even, where the matrix elements
\[ < {\cal P}_{2k}^0(\Omega_2) | {\cal P}_{2k}^0(\Omega_1) > = (-1)^k \]
and
\[ < {\cal P}_{2k}^0(\Omega_i) | {\cal P}_{2k}^0(\Omega_{12}) >
 = \left\{ \ba{ll} 0 & k\ \mbox{odd} \\ \frac{(-1)^{k/2}}{k+1} & k\
 \mbox{even} \ea \right. \]
are used in agreement with equation (25) of reference\cite{fab}.

For triplet $S$-states the $B_k(\Omega)$ for $k$ odd are suitable for
the motion of the two electrons around the nucleus, but
 ${\cal P}_{2k}^0 (\Omega_{12})$
 is not appropriate since it is even in the exchange of the
two electrons. In order to find the second set of HH we note that, since
the normalized PH are proportional to Gegenbauer polynomials (\ref{21}), then
\[ < {\cal P}_{2k}^0(\Omega,\varphi) | {\cal P}_{2k'}^0
(\Omega,\varphi) > = \delta_{k,k'}\;. \]
By taking the derivative with respect to $\varphi$, noting
\[ < \frac{\partial {\cal P}_{2k}^0}{\partial \varphi} | 
{\cal P}_{2k'}^0 > = 0\,, \]
one generates the additional  basis elements
\[ \frac{\partial {\cal P}_{2k}^0(\Omega,\varphi)}{\partial \varphi} =
\frac{2}{\pi^{3/2}} \left[ x\sqrt{1-z^2}\,\cos 2\varphi -
 z\,\sin 2\varphi\right]\;C_{k-1}^2\left( z \cos 2 \varphi +
 x\sqrt{1-z^2}\,\sin 2 \varphi \right)\,. \]

 For $\varphi=0$ one obtains the set of HH where electrons in a
$p$-state are coupled to form an $S$-state obviously orthogonal to
$B_k(\Omega)$, where both electrons are in an $s$-state.

 For $\varphi = 3\pi/4$, one obtains the required basis
$z C_k^{2}(x\sqrt{1-z^2})$, which however must be normalized and has to
be made orthogonal to $B_k(\Omega)$ for $k$ odd. The desired basis
elements are

\[ B_k^\bot(\Omega) = \left\{ \ba{ll}
\frac{2\sqrt{3}}{\pi^{3/2}\,\sqrt{k(k+2)}} z C_{k-1}^2(x\sqrt{1-z^2}) &
 k > 0\ \mbox{even} \\ 
\frac{2\sqrt{3}}{\pi^{3/2}\,\sqrt{k(k+2)-3}} \left[ z C_{k-1}^2(x\sqrt{1-z^2}) -
\frac{(-1)^{(k-1)/2}}{2} C_k^1(z) \right] & k > 1 \ \mbox{odd} \ea \right.
 \]

It is worthwhile to note that, since $z(3\pi/4) = r_{12}^2/r^2 - 1$,
the basis element $B_k^\bot$ describes the electron pair in an
$S$-state. Then, in the expansion
\[ \Psi = r^{-5/2} \sum_{k,k_{\bot}} \left( B_k u_k(r) +
B_{k_{\bot}}^\bot u_{k_{\bot}}^\bot(r)\right) \]
the electrons are individually in $s$-states in $B_k$. (Here and in the
appendix, we sometimes emphasize that indices take on the values that
are associated with the perpendicular set by using the notation $k_{\bot}$
or $k'_{\bot}$. Where there is no room for confusion, we drop the $\bot$.)
This means
that in our approximation the space wave function becomes the sum of
two functions. In the singlet state
\[ \Psi_S = \psi_S(r_2^2 - r_1^2, r) + \psi_S^\bot(r_{12}^2, r) \]
where $\psi_S$ and $\psi_S^\bot$ are even in an exchange of the two
electrons for the singlet state, while in the triplet state
\[ \Psi_T = \psi_T(r_2^2 - r_1^2, r) + (r_2^2-r_1^2) \psi_T^\bot(r_{12}^2, r)
\]
where $\psi_T$ is odd in the exchange of the two electrons.

 The partial waves are solutions of the system of coupled second order
differential equations
\be
(T_{k} - E) u_{k} + \sum_{k',k'_{\bot}} (U_{k}^{k'}u_{k'} + U_{k}^{k'_{\bot}}
u_{k'_{\bot}}^\bot) = 0  \label{22} \ee
and
\be (T_{k_{\bot}} - E) u_{k_{\bot}}^\bot + \sum_{k',k'_{\bot}} (U_{k_{\bot}}
^{k'_{\bot}}  u_{k'_{\bot}}^\bot + U_{k}^{k'} u_{k'}) = 0 \label{23} \ee
where
\[ T_k =  - \frac{d^2}{dr^2} + \frac{{\cal L}_k({\cal L}_k+1)}{r^2} \]
and ${\cal L}_k= 2k + 3/2$, (both valid for $k$ and $k_{\bot}$).
\esp

The matrix elements of the potential matrix are given in Appendix A.

\newpage
\begin{center}
\section*{Results and Analysis}
\end{center}

The $r$ dependence of the coupling matrix is simply $1/r$ (i.e.
multiplying a matrix of constant coefficients). To solve the resulting
coupled differential equations, we take advantage of a method developed
by V. Mandelzweig and M. Haftel\cite{Man} - and embodied in one of their
numerical
programs. It was used by Haftel and us (here, Fabre and Larsen) to
obtain results for the singlet\cite{haf}. Michael kindly made available
to us his program, that we used for the triplet.

Its essence is to calculate the wavefunction and its logarithmic derivative
by using expansions - and, then, obtain the matrices of coefficients
(of these expansions) by numerical recursions.

We solve the set of equations for a number of values of $k_{max}$, such
as $33, 37, 41$ ... and extrapolate the result to infinite order.
We can compare our results to a wealth of reliable data\cite{Frank,addit}.
\vspace{.4cm}
\subsubsection*{Singlet}
As noted above, the results have been published\cite{haf}. The result for the
ground state is that we obtained an extrapolated value of $2.903 471$
atomic units,
compared to the more accurate variational result of $2.903 724$.
(We truncate all our numbers to 6 decimals.) The error was thus $\sim .01\%$.

For the excited state, the extrapolated value was $2.145 825$
compared with a variational value of $2.145974$. The resulting
error was thus $\sim .007\%$.

These were based on calculations for $k_{max} = 24,28,32,36$ and 
noting\cite{schnei} that, asymptotically, the differences between the energies
for successively larger values of $k_{max}$ decrease at least as
fast as $k_{max}^{-4}$.
The differences were then fitted to an inverse polynomial:
\bdm
\Delta{E(k)} = E(k) - E(k-4) = \frac{1}{k^2(A + B\,k + C\,k^2)}
\edm
and the extrapolated values were obtained by adding the
remaining contributions that would result from summing to infinity.

Another point has, in retrospect, proved important. For the ground state,
the value for $k_{max} = 36$ is $2.903 067$, i.e. this `raw'
results is already correct to 4 digits. The extrapolated `tail' only adds
a modest amount. For the excited state, the situation is quite different.
The `raw' value is $2.129 457$  and we see that therefore the 
difference between the resulting energy and the correct value is,
roughly, $1 \%$.  The contribution of the extrapolated `tail' is very
important yielding several more digits in agreement.
\subsubsection*{Triplet}
In Table $1$, we present the numerical results (in a.u.) obtained by solving
the coupled equations. These need, again, to be extrapolated, so as to take
into  account the contribution of the higher order potential harmonics, and
then compared with good variational estimates: $2.1752495$ for the lowest
state, $2.068685$ for the excited state. Without the $r_{12}$ interaction, the 
exact result is $2.5$ a.u. 
\begin{table}[h]
\caption{energies from solving the equations up to $k_{max}$}
\begin{tabular}{llll} \hline \hline
$k_{max}$ & ground state  & excited state  & no $r_{12}$ interaction \\ \hline
   21 &  2.1401236235411021    &    1.9201801253586101 &   \\
   23 &  2.1459023327465288    &    1.9377802892305477 &   \\
   25 &  2.1504882412407837    &    1.9525548534156408 &   \\
   27 &  2.1541746943943439    &    1.9650792663836013 &   \\ 
   29 &  2.1571714798190449    &    1.9757878236977571 &   \\ 
   31 &  2.1596321992027180    &    1.9850138971543965 &   \\  
   33 &  2.1616708453178672    &    1.9930169776811079 &   \\  
   35 &  2.1633735780590255    &    2.0000017724075086 &   \\
   37 &  2.1648062043495217    &    2.0061315935556941 &   \\  
   39 &  2.1660197311520152    &    2.0115381473492293 &   \\ 
   41 &  2.1670540234821436    &    2.0163286338520296 &   \\
   43 &  2.1679406196279248    &    2.0205910881663552 &   \\ 
   45 &  2.1687046390433931    &    2.0243983641445255 &   \\  
   47 &  2.1693662957017716    &    2.0278112036995569 &  2.4978184400838062 \\
   49 &  2.1699419469276808    &    2.0308805802207102 &  2.4980527934858051 \\
   51 &  2.1704449472143039    &    2.0336495409553854 &  2.4982549671085871 \\
   53 &  2.1708862505879244    &    2.0361546418675421 &  2.4984302656593832 \\
   55 &  2.1712749122742111    &    2.0384270954208123 &  2.4985829842420010 \\
   57 &  2.1716184476775823    &    2.0404936798376949 &  2.4987166244132616 \\
   59 &  2.1719231375119347    &    2.0423774774219713 &  2.4988340592267531 \\
   61 &  2.1721942486212956    &    2.0440984689458583 &  2.4989376603658377 \\
   63 &                        &    2.0456740179909737 &  2.4990293968470676 \\
   65 &                        &    2.0471192766548691 &  2.4991109122218091 \\
   67 &                        &    2.0484475085284343 &  2.4991835853845747 \\
   69 &     &            &  2.4992485787876214  \\ \hline  \hline
\end{tabular}
\end{table}
\newpage

The first thing that we notice, then, is that, even for the much higher 
values of $k_{max}$ that we have used for the triplet, the difference 
between our results and the variational ones is still more than $1$ part in a 
thousand for the ground state and $1\%$ for the higher state.
If we had limited ourselves to the orders used in the singlet, our errors
would have been $1/2 \%$ and $3 \%$, respectively!

We were thus driven to higher values of $k_{max}$ and thus to a greater number
of states in the calculations. For example, a $k_{max}$ of $61$ corresponds
to a total number of states of $91$. Even then, we also had to work harder
at the extrapolation than we did for the singlet.
The differences were fitted to an inverse polynomial with more coefficients:
\bdm
\Delta{E(k)} =  \frac{1}{A\,k^4 + B\,k^3 + C\,k^2 + D\,k + E}
\edm
with the $\Delta$s variously defined as
$\Delta{E(k)} = E(k) - E(k-4)$ and $\Delta{E(k)} = E(k) - E(k-2)$
and the extrapolated values obtained by using this fit, for the differences,
to obtain the higher order contributions.


Examination of the results of the extrapolations then shows that the best
strategy consists in fitting the coefficients with the minimum of data points,
taken at the highest values of $k_{max}$.
For the ground state, the use of $\Delta(k) = 4$ yields the best result.
For the excited state, $\Delta(k) = 2$, is superior.

Thus we find that for the ground state, the use of the set of values of 
$k_{max}$:  (51,53,55,57,59,61) yields  $2.175042$, while 
          (41,45,49,53,57,61) gives us our best result $2.175144$ (both
results, here, rounded to 6 decimal places). We recall that this must be
compared with $2.1752495$.

For the excited case, (55,57,59,61,63,67) gives  $2.068704$, while
(47,51,55,59,63,67) gives  $2.068757$, also rounded at the same place.
We compare this with $2.068685$.

Since in our method we only approximate by truncating the basis, both the
elements which appear in it and the highest order that we take into
account in our calculations, our results must obey the variational principle
and must yield upper bounds in the (negative) energies of the states.
We can verify that this is indeed the case for our "raw" results, displayed
in Table $1$. For the extrapolated results, this is only evident for the 
ground state. Our results for the excited state are slightly deeper than the
correct energy.

This is an artifact of the extrapolation. Clearly, our inverse polynomial 
only imperfectly represents the correct asymptotic behaviour of the 
differences. In fact taking different sets of points, from which to
extrapolate, and trying to `see' trend lines, we think that we see oscillations
and certainly instability and serious limitations in our ability to
extrapolate.

Nevertheless, we obtain energies to within $1$ or $2$ parts in $20,000$ for
the lowest state and $1$ to a few parts in $100,000$ for the excited state.

Finally, we note an extrapolation with $\Delta(k)$ of $2$, for the case
where the electrons are uncoupled. Our result is then 2.499999995, rounded
to nine decimal places, compared to the exact value of $2.5$\, In this case
our basis is complete, aside from truncation in the order, and we take
this result as a confirmation of the soundness of our numerical approach
and procedures.

\newpage
\begin{center}
\section*{Conclusion}
\end{center}

We believe that we have demonstrated that the use of Potential Harmonics
is a sound way of obtaining reliable results, at least for 2-electron atoms,
to an accuracy of $1$ part in $10^4$ or $10^5$.
Our aim was to confirm this, with our first two calculations, and then to 
proceed with higher excited states for these 2-electron atoms and, perhaps
more excitingly, with 3-electron atoms!

While this is probably feasible, and in fact the potential harmonics seem
to yield better accuracy for the excited states as 3-body and higher 
correlations become less important, our experience leads us to the following
caveat.

We have found many aspects of our effort, such as the calculation of our 
matrix elements, the necessary extrapolations, to be quite delicate!
We would prefer other methods which, using integral equations, 
would not require them.

Our final and best thought, then, is to draw attention to the simplicity
of the form of the wave functions that we have used, and to the fact that,
in them, the pairs are, each, in an $s$-state.
\vspace{1in}
\begin{center}
{ \bf Acknowledgment}
\end{center}
Sigurd Larsen gratefully acknowledges a sabbatical from Temple University 
for the year $96$-$97$ and the hospitality of the Physics department of
the University of Pretoria, S.A.
as well as of the School of Computing and Mathematics of Deakin University,
Geelong, Victoria, Australia.

\newpage
\begin{center}
\section*{Appendix A}
\end{center}
The matrix elements in equations (\ref{22}) and (\ref{23}),
 after making the potential harmonics basis substitution,
 take the form of
integrals of products of three Gegenbauer polynomials.
 Let us introduce a number
of expressions for sums that occur from particular combinations of
HHs and PHs:
\[ S_k^{k'} = \sum_{\ba{c} {\scriptstyle m = |k-k'|} \\ 
{\scriptstyle \rm{(step\ 2)}} \ea}^{k+k'}
 \frac{(m+1)}{(2m+1)(2m+3)}\,, \] 
 where $S_{k'}^k = S_k^{k'}$,
\[ \sigma_k^{k'} = \sum_{\ba{c} {\scriptstyle m=|k-k'|} \\ 
{\scriptstyle \rm{(step\ 2)}} \ea}^{k+k'}
 \frac{(-1)^{m/2}}{(2m+1)(2m+3)}\,, \]
 and $\sigma_{k'}^k = \sigma_k^{k'}$.
\begin{eqnarray*} \lefteqn{{\displaystyle \sum}_k^{k'} =} \\ 
 & & \sum_{\ba{c} {\scriptstyle m=|k-k'|+1} \\ {\scriptstyle 
 \rm{(step\ 2)}} \ea}^{k+k'+1}
(m+1) \frac{(k+k'-m+1)(k-k'+m+1)}{(2m+5)(2m+3)} \times \\
& & \frac{(k'-k+m+1)(k+k'+m+3)}{(2m+1)(2m-1)}\,,
 \end{eqnarray*}
with $ \Sigma_{k'}^k = \Sigma_k^{k'}$.
\begin{eqnarray*} \lefteqn{{\displaystyle (\sum}^{\bot})_k^{k'} =} \\
  & & \frac{1}{16} \sum_{\ba{c}
 {\scriptstyle n = |k-k'|} \\ {\scriptstyle
 \rm{(step\ 2)}} \ea}^{k+k'} (-1)^{n/2} \frac{(n+1)}{(2n+1)(2n+3)}
 \sum_{\ba{c} {\scriptstyle m=|k-k'|+1} \\ {\scriptstyle 
\rm{(step\ 2)}} \ea}^{k+k'+1}
 (\delta_{m,n+1} - \delta_{m,n-1}) \times \\
 & & \frac{(m-k+k'+1)(m+k-k'+1)(k+k'+1-m)(k+k'+3+m)}{m(m+2)}\,,
\end{eqnarray*}
 where, by symmetry, $(\Sigma^\bot)_{k'}^{k} = (\Sigma^\bot)_k^{k'}$.
\ \par

\esp
\esp
\subsection*{The matrix elements}

First, consider the contribution from the two electrons
separately around the nucleus, where $k$ and $k'$ are both odd because
of the anti-symmetric requirement of the triplet wave function.
\begin{eqnarray}
U_k^{k'} & = & \frac{2}{r} \left< B_{k} \mid \frac{1}{r_{12}}
 \mid B_{k'} \right> - \frac{2}{r} \left< B_{k}\mid
\frac{Z}{r_1} + \frac{Z}{r_2} \mid B_{k'} \right> \nonumber \\
 & = & \frac{64}{\pi r} \left[ \frac{\sqrt{2}}{4} \sigma_k^{k'} 
 - Z S_k^{k'} \right]\,, \label{100}
\end{eqnarray}
\noindent Then, consider the interactions between the electron-electron
wave function and the electron-nucleus wavefunctions. Here, we can
have $k$ odd and $k' \ne 0$ even,
\begin{eqnarray}
U_k^{k'_{\bot}} & = & \frac{2}{r} \left< B_{k} \mid \frac{1}{r_{12}}
 \mid B_{k'}^\bot \right> - \frac{2}{r} \left< B_{k}
 \mid
 \frac{Z}{r_1} + \frac{Z}{r_2} \mid B_{k'}^\bot \right> \nonumber \\
 & = & \frac{(-1)^{(k-1)/2} 16 \sqrt{6}}{\pi r \sqrt{k'(k'+2)}\; k(k+2)}
 {\displaystyle \sum}_k^{k'}\,, \label{101}
\end{eqnarray}
 and for the case $k'$ odd and $k \ne 0$ even, we use 
$U_{k'}^{k_{\bot}} = U_{k_{\bot}}^{k'}$.

\noindent For the case where $k$ and $k'$ are both odd,
\begin{eqnarray}
 U_k^{k'_{\bot}} & = & \frac{16 \sqrt{6}}{\pi r \sqrt{k'(k'+2)-3}} \left[
 \frac{(-1)^{(k-1)/2}}{k(k+2)} {\displaystyle \sum}_k^{k'} - (-1)^{(k'-1)/2}
 \sigma_k^{k'} \right]\,. \label{102} 
\end{eqnarray}

\noindent
Finally, we need to consider the perp-perp matrix elements.
Thus, we have for $k$ and $k'$ both even:
\begin{eqnarray}
 U_{k_{\bot}}^{k'_{\bot}} & = & \frac{2}{r} \left< B_{k}^\bot \mid
 \frac{1}{r_{12}} \mid B_{k'}^\bot \right> - \frac{2}{r} \left<
 B_{k}^\bot \mid \frac{Z}{r_1} + \frac{Z}{r_2} \mid
 B_{k'}^\bot \right> \nonumber \\
 & = & \frac{16}{\pi r \sqrt{k(k+2)}\;\sqrt{k'(k'+2)}}\left[
 \sqrt{2}\;{\displaystyle \sum}_k^{k'} - 12\; Z
({\displaystyle \sum}^\bot)_k^{k'}\right]\,, \label{103}
\end{eqnarray}
 $k$ and $k'$ both odd,
\begin{eqnarray}
 U_{k_{\bot}}^{k'_{\bot}} & = & \frac{16}{\pi r
\sqrt{k(k+2)-3}\;\sqrt{k'(k'+2)-3}} \times \nonumber \\
  & & \left\{\sqrt{2}\;\left[ \left( 1 - \frac{3}{k(k+2)} -
 \frac{3}{k'(k'+2)} \right)
 {\displaystyle \sum}_k^{k'} + 3(-1)^{|k-k'|/2}\sigma_{k'}^k
 \right] \right. \label{104} \\
 & - & \left. 12\; Z \left[ ({\displaystyle \sum}^{\bot})_k^{k'} - 
(-1)^{(k+k')/2}
S_k^{k'} \right] \right\} \,, \nonumber
\end{eqnarray}
$k$ odd and $k'$ even,
\begin{eqnarray}
 U_{k_{\bot}}^{k'_{\bot}} & = & \frac{16 \sqrt{2}}{\pi r \sqrt{k(k+2)}
 \sqrt{k'(k'+2)-3}} \left( 1 - \frac{3}{k'(k'+2)}\right) {\displaystyle
\sum}_k^{k'}\,, \label{105}
\end{eqnarray}
 and finally, with $k$ even and $k'$ odd,
\begin{eqnarray}
 U_{k_{\bot}}^{k'_{\bot}} & = & \frac{16 \sqrt{2}}{\pi r \sqrt{k(k+2)}
 \sqrt{k'(k'+2)-3}} \left( 1 - \frac{3}{k'(k'+2)} \right) {\displaystyle
\sum}_{k'}^k\,. \label{106}
\end{eqnarray}

\end{document}